\documentclass[conference]{IEEEtran}
\IEEEoverridecommandlockouts
\usepackage{cite}
\usepackage{amsmath,amssymb,amsfonts}
\usepackage{algorithmic}
\usepackage{graphicx}
\usepackage{textcomp}
\usepackage{xcolor}
\usepackage{cite}
\usepackage{balance}
\usepackage{url}
\usepackage{booktabs}
\usepackage{multirow}
\usepackage{multicol}
\usepackage{subcaption}

\usepackage{etoolbox}
\makeatletter
\patchcmd{\@makecaption}
  {\scshape}
  {}
  {}
  {}
\makeatother

\usepackage{multirow}
\def\BibTeX{{\rm B\kern-.05em{\sc i\kern-.025em b}\kern-.08em
    T\kern-.1667em\lower.7ex\hbox{E}\kern-.125emX}}

\begin{document}
\bstctlcite{IEEEexample:BSTcontrol}
\newcommand{\sid}[1]{\color{red}{ #1 } \color{black}}
\newcommand{\outline}[1]{\color{gray}{ #1 } \color{black}}
\newcommand{\fs}{$\texttt{fp16}$}

\title{Sustainable Supercomputing for AI: GPU Power Capping at HPC Scale}

\author{Dan Zhao\IEEEauthorrefmark{2}\textsuperscript{\textsection},
Siddharth Samsi\IEEEauthorrefmark{1}\textsuperscript{\textsection}, 
Joseph McDonald\IEEEauthorrefmark{1},
Baolin Li\IEEEauthorrefmark{3},\\
David Bestor\IEEEauthorrefmark{1},
Michael Jones\IEEEauthorrefmark{1}, 
Devesh Tiwari\IEEEauthorrefmark{3},
Vijay Gadepally\IEEEauthorrefmark{1}\\
\IEEEauthorrefmark{1} MIT, 
\IEEEauthorrefmark{2} NYU,
\IEEEauthorrefmark{3} Northeastern University
}

\maketitle

\begingroup\renewcommand\thefootnote{\textsection}
\footnotetext{Corresponding authors : dz1158@nyu.edu, samsi.1@osu.edu}
\endgroup

\begin{abstract}
As research and deployment of AI grows, the computational burden to support and sustain its progress inevitably does too. To train or fine-tune state-of-the-art models in NLP, computer vision, etc., some form of AI hardware acceleration is virtually a requirement. Recent large language models require considerable resources to train and deploy, resulting in significant energy usage, potential carbon emissions, and massive demand for GPUs and other hardware accelerators. However, this surge carries large implications for energy sustainability at the HPC/datacenter level. In this paper, we study the aggregate effect of power-capping GPUs on GPU temperature and power draw at a research supercomputing center. With the right amount of power-capping, we show significant decreases in both temperature and power draw, reducing power consumption and potentially improving hardware life-span with minimal impact on job performance. While power-capping reduces power draw by design, the aggregate system-wide effect on overall energy consumption is less clear; for instance, if users notice job performance degradation from GPU power-caps, they may request additional GPU-jobs to compensate, negating any energy savings or even worsening energy consumption. To our knowledge, our work is the first to conduct and make available a detailed analysis of the effects of GPU power-capping at the supercomputing scale. We hope our work will inspire HPCs/datacenters to further explore, evaluate, and communicate the impact of power-capping AI hardware accelerators for more sustainable AI.
\end{abstract}

\section{Introduction}
Recent advances in the capabilities AI systems have delivered stunning achievements ranging from realistic natural language generation to facilitating drug discovery and winning complex, strategic games---but not without cost. For example, training large-scale NLP models can emit CO$_2$ amounts commensurate with the lifetime emission of up to five cars \cite{strubell2019}. Many of these models are commonly characterized by having hundreds of millions of parameters or even more (e.g., GPT-4, Stable Diffusion) sometimes requiring weeks to months \cite{megatron2021,workshop2023bloom,shoeybi2020megatronlm,llama2} of training on specialized hardware and petabytes of data to realize such impressive performance. At the same time, these large-scale models are also seeing widespread adoption and application from commercial usage to non-traditional domains such as information retrieval and drug discovery. If current trends continue, these models will require ever-growing resources in the form of training data, compute resources, time, and energy. 

Along with training on large datasets, deploying models can also require significant energy consumption. Models like GPT-3/4 that power applications such as ChatGPT and existing search engines can be a massive source of energy use as millions of users use these tools daily. While large language models (LLMs) are currently attracting significant amounts of attention, considerable resources are also used to power training and inference of various machine/deep learning systems in other domains with applications in object recognition, recommendation engines, autonomous driving, bio-informatics, etc. The race towards realizing further advances in AI has also spurred many large, key players to construct new AI research supercomputing centers \cite{metaAI} with massive amounts of compute and storage geared towards enabling ever larger computational workloads for AI research. The overall AI research ecosystem itself can also induce a considerable carbon footprint aside from compute in the form of travel (to/from conferences), concentration of conference deadlines in warmer months, and more \cite{greenworldAI}.

At the core of these intensive resource needs is the issue of sustainability: these systems will not only continue to grow in complexity and dimensionality to achieve increasingly impressive milestones, but they will also see increasing proliferation and adoption across various domains. These factors can increasingly impact the climate, stressing the sustainability of energy sources, as well as other resources, as the power needs and carbon emissions of these systems grow. 
Therefore, finding ways to enable optimal trade-offs between performance, efficiency, and sustainability becomes a question of first-order importance for the climate and the sustainability of continued AI development. Substantial progress has been made in improving model efficiency through techniques like model pruning \cite{xia-etal-2022-pruning}, model distillation \cite{hinton2015distilling}, model sparsification \cite{sparsity}, quantization \cite{coelho2021automatic},  and data-centric approaches that can enable model training on smaller representative datasets \cite{mansheej2021}. However, many of these approaches rely primarily on a reduction in the size of the model or data which can require specialized knowledge or significant modifications to the training/deployment process. 

One potential path for academic HPC and commercial cloud datacenter operators to reduce AI's energy consumption at scale is to leverage the ability to limit the power drawn by the hardware itself. In this vision paper, we present empirical observations and a statistical analysis of the effect of power capping GPUs at an academic supercomputing center where we deployed a 60\% power cap on GPUs. We show that limiting the power used by GPUs across the datacenter has the potential to reduce GPU energy consumption at the datacenter while also reducing operating temperatures. We argue that power-capping should be adopted in data centers to help extend the lifetime of hardware, amortize the embodied carbon of data centers, and more. 

\section{Prior Work}\label{sec:prior}
The idea of power limiting hardware to realize energy efficiency is not new and has been studied extensively. However, to the best of our knowledge a significant amount of previous work on studying the effect of power capping on hardware performance has focused primarily on CPUs. For example, \cite{dongarra} analyzed the effect of power caps on numerical applications and \cite{Yongyan} developed a power capping architecture for controlling the power consumption of large scale clusters. \cite{feng-sc05} developed a power-aware algorithm to change CPU voltage and frequency to reduce power consumption while minimizing the impact on compute performance. \cite{tang2016power} showed that power capping resulted not only in lower hardware operating temperatures but also had the effect of increasing the mean time between failures. As GPUs have become widely deployed for AI workloads, there has been a renewed interest in studying the effect of power capping on GPU performance, specifically with the goal of quantifying energy savings. 

For example, \cite{mcdonald2022} showed that pre-training the BERT \cite{devlin-etal-2019-bert} model at 150W or 60\% of peak power required 8.5\% additional compute time but saved 12.5\% in energy consumed. Similarly, \cite{frey2022} studied the performance of natural language models, computer vision models and graph neural networks and showed that power limiting GPUs for AI workloads offers an effective way for saving energy. Finally, \cite{jerzy-gpu-powercap2023} showed energy savings of at least 22\% on two commonly used NVIDIA GPU accelerators and also developed an optimizer for such workloads. 
Meanwhile, other institutions have also experimented with power capping GPUs in efforts to improve the energy efficiency of their supercomputing resources; the Flatiron Institute \cite{flatiron} has experimented with lowering the peak power GPUs of one of their clusters from 400W per GPU to 225W, saving energy while preserving 90\% of performance. However, these larger-scale experiments with power caps and their effects, if performed, have not always been made publicly available or received rigorous statistical analysis of the results. Others \cite{iccs2022} have examined the effects of power-capping on a more limited scale, restricting their attention to the effect of power-capping a single or handful of GPUs on the efficiency and performance of a few deep vision models. Finally, other works \cite{hotcarbon, noman_sustain} have taken a broader scope, examining themes such as flexible optimization of datacenters in the face of climate change, renewable energy opportunities, and the future implications of continued trends in compute/carbon optimization.

\section{Data \& Methodology}
We conducted our experiments on the MIT Supercloud high-performance computing (HPC) system \cite{reuther2018interactive}: a 7-petaflop heterogeneous system that consists of 224 Intel Xeon Gold 6284 nodes each with two NVIDIA Volta V100 GPUs with 32 GB of RAM and 384 GB of system memory. Each node on the system has two independent back-end fabrics: a 100 Gb/s Intel Omnipath and a 25 Gb/s Ethernet interconnect using Mellanox ConnectX-4 adapters with all servers connected to a single, non-blocking Arista DCS-7516 Ethernet core switch. The GPUs, Omnipath, and Ethernet cards are all connected to PCIe slots that route directly to the Xeon processors without any intermediary PCIe switches. The system also has 480 Intel Xeon Platinum 8260 nodes with 194 GB of RAM. This system uses the Slurm scheduler for resource management. Data was collected via Slurm for transparent, lightweight and automated system monitoring. The scheduler provides the ability to use a job prologue to start data collection and a job epilogue to terminate data collection at the end of a job. All experiments exclusively used the 25 Gb/s Ethernet interconnect and run exclusively on NVIDIA GPUs. 

For each job that requested a GPU, we used the \texttt{nvidia-smi} utility to collect fine-grained GPU utilization data on every GPU allocated to a job. This monitoring data was collected on a 100 ms interval. The data sample used for this analysis was collected at a point shortly before system-wide implementation of power caps, consisting of a total of 123,204 GPU jobs of which 19,676 were subject to power-caps and the remaining were not. Features include the job's run-time and the minimum, maximum, variance, average, and percentiles of the job's GPU utilization, GPU temperatures, power draw, etc. over the job's run time. We summarize the data used in our analysis for some of the most relevant variables in Table \ref{tbl:summ_stats}. Since each job is itself a multi-variate time-series of different features relating to energy, utilization, etc., the statistics in Table \ref{tbl:summ_stats} are taken over the entirety of the job's runtime time series.

Our analysis focuses primarily on the job-level hardware utilization due to user privacy concerns. The HPC cluster consists primarily of V100 GPUs and a very limited number of A100s. Given the shared nature of the cluster and its heavy utilization, it is difficult to vary power-cap levels too frequently due to disruption risks for users of the system, limiting our ability to systematically experiment with a wider range of power-capping levels across all GPUs. As users of the HPC cluster can run highly diverse workloads, detailed collection of data on specific jobs can also be difficult given low user response-rates to surveys and our desire to protect user privacy.

\begin{table}[htbp!]
\small
\begin{tabular}{lrrrr}
\toprule
Metrics &  Time (minutes) &  Util. (\%) & Temp. (C)&  Power (W) \\
\midrule
 mean &          595.08 &             31.79 &             39.66 &               69.58 \\
  s.d. &         1794.42 &             31.81 &             10.14 &               43.75 \\
  min &            1.02 &              0.00 &             18.95 &               21.89 \\
  25\% &           20.58 &              0.26 &             32.00 &               34.05 \\
  50\% &          145.36 &             24.84 &             37.43 &               53.99 \\
  75\% &          610.42 &             55.88 &             45.12 &               94.70 \\
  max &        65023.47 &            100.00 &             81.99 &              237.33 \\
\bottomrule
\end{tabular}
\caption{\small Summary statistics of selected features/variables relevant to all the jobs analyzed in this paper. ``s.d.'' denotes the sample standard deviation while 25\% / 50\% / 75\% denote the 25th, 50th, and 75th percentiles, respectively. ``Time'' refers to job runtime in minutes, ``Util.'' refers to GPU utilization as a percentage, ``Temp.'' to temperature in Celsius, and ``Power'' to power draw in Watts.}
\label{tbl:summ_stats}
\end{table}

\section{Results}\label{sec:experiments}
With power caps implemented on a system-wide level, we empirically observed improvements in energy usage especially via reduced power draw and temperatures of these GPUs. At the same time, we found minimal impact on job performance under these caps in a way consistent with others \cite{flatiron, frey2022, mcdonald2022}. While power-capping naturally reduces power draw by design, the aggregate effect system-wide in the context of overall energy consumption is less clear; for instance, if users notice job performance degradation from GPU power-caps, they may request additional GPU-jobs to compensate, negating any energy savings from power-capping in the first place (or even worsening energy consumption).

As such, a more rigorous treatment and analysis of these effects on power draw and temperatures across jobs can help determine how effective power capping is to a degree of statistical certainty. Doing so can help inform whether power capping produced significant changes, different from random chance, in improving energy sustainability and whether the magnitude of said change is sufficient to be worth such an intervention (i.e., implementing power caps) in the first place.

\subsection{An Empirical Analysis of Power Capping}
\begin{figure}[htbp!]
    \centering
    \includegraphics[width=0.4\textwidth]{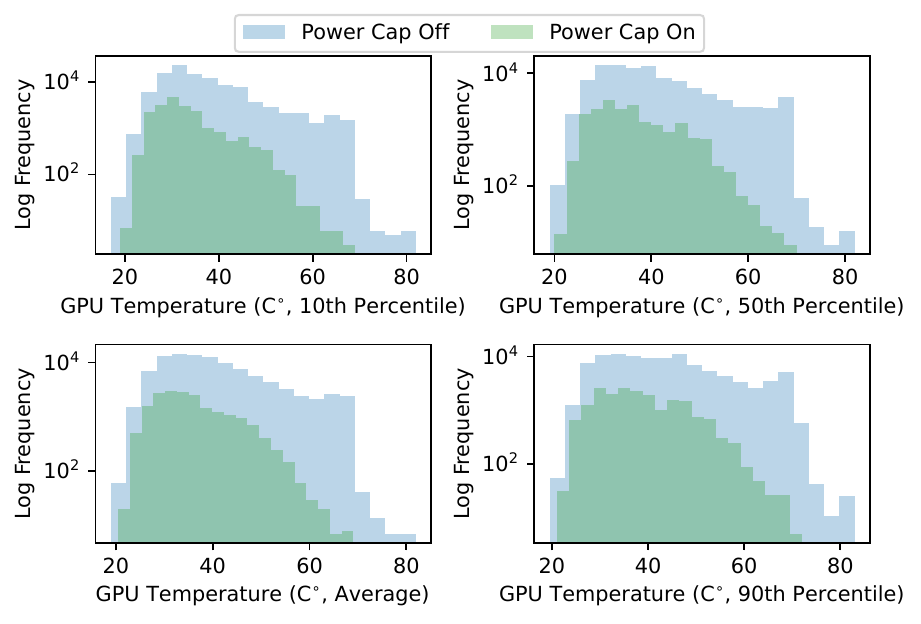}
    \caption{\small Distribution of GPU temperatures (Celsius) from jobs with and without power capping. Shown here are the empirical distribution of average temperatures as well as distributions of temperatures of the 10th, 50th, and 90th percentiles across individual jobs with and without power caps. Note that the y-axis for frequency is on log scale.}
    \label{figure:gpu_temp_hist}
\end{figure}

\begin{figure}[htbp!]
    \centering
    \includegraphics[width=0.4\textwidth]{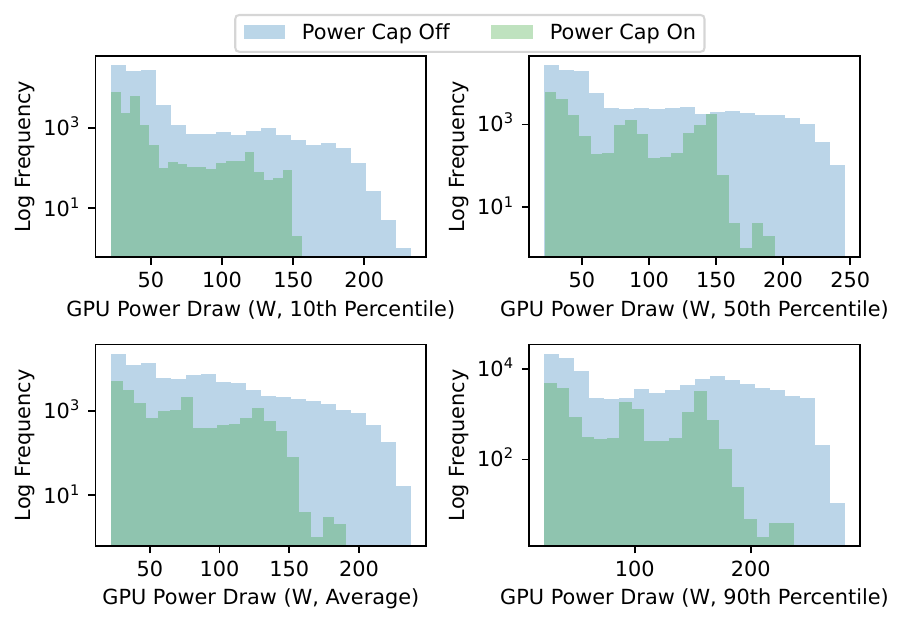}
    \caption{\small \textbf{Distribution of GPU power draw (Watts) from jobs with and without power capping.} Shown here are the distribution of average temperatures as well as distributions of temperatures of the 10th, 50th, and 90th percentiles across individual jobs with and without power caps.}
    \label{fig:gpu_powerdraw_hist}
\end{figure}

Figure \ref{figure:gpu_temp_hist} and Figure \ref{fig:gpu_powerdraw_hist} provide a bird-eye's view of our empirical observations on GPU temperatures and power draw between power-capped and un-capped jobs. Figure \ref{figure:gpu_temp_hist} shows the distributions of GPU temperatures from jobs with and without GPU power caps across a range of percentiles. The underlying data contains statistics of individual jobs that track features throughout each job's run; for instance, these include the highest and lowest GPU temperatures and power draw of each job, average and percentiles of GPU utilization rates of each job, etc.

\begin{figure}[htbp!]
    \centering
    \includegraphics[width=0.5\textwidth]{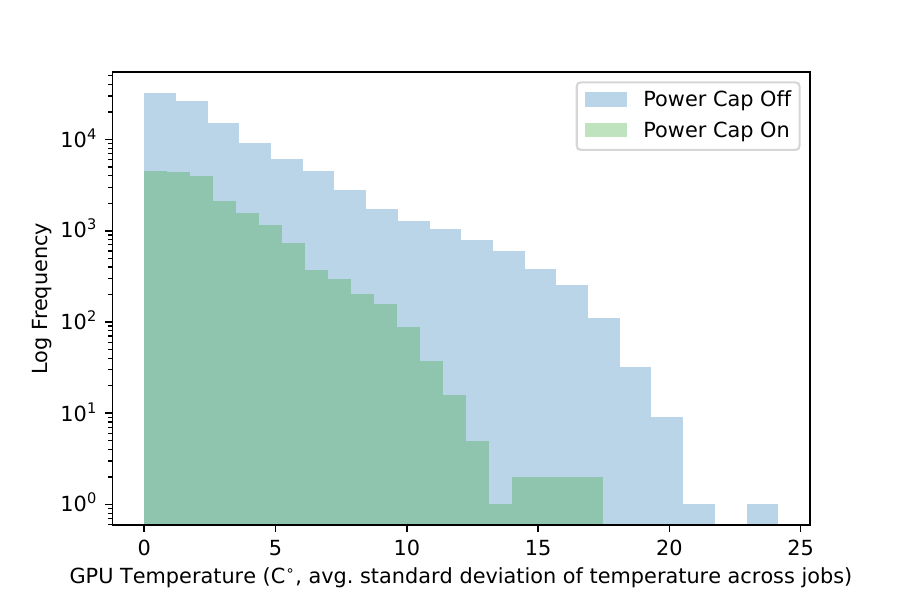}
    \caption{\small \textbf{Distribution of average standard deviation of GPU temperatures from jobs with and without power capping.} Shown here are the distribution of average standard deviation of GPU temperatures of individual jobs with and without power caps. We note that power-capped jobs show a smaller and more stable range of temperature fluctuations than uncapped jobs.}
    \label{fig:gpu_temp_histsd}
\end{figure}

\begin{figure}[htbp!]
    \centering
    \includegraphics[width=0.5\textwidth]{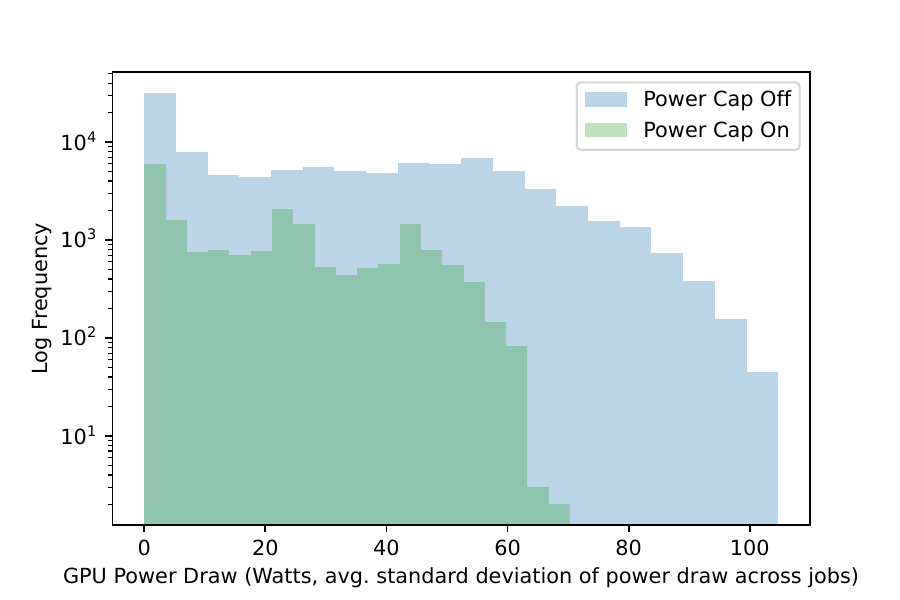}
    \caption{\small \textbf{Distribution of average standard deviation of GPU power draw from jobs with and without power capping.} Shown here are the distribution of average standard deviation of GPU temperatures of individual jobs with and without power caps.}
    \label{fig:gpu_powerdraw_histsd}
\end{figure}

In Figure \ref{figure:gpu_temp_hist}, we see that the power-capped jobs have lower GPU temperatures than those on un-capped jobs across the whole distribution---both lower on average temperatures as well as lower temperatures across the 10th, 50th, and 90th percentiles. We also see from Figure \ref{fig:gpu_temp_histsd} that the variance or standard deviation of GPU temperatures, on average, is also lower on power capped jobs when compared against un-capped ones. In Figure \ref{fig:gpu_powerdraw_hist}, we see a similar trend for GPU power draw which we expect---power capping decreases power draw across the distribution. However, we also see in Figure \ref{fig:gpu_powerdraw_histsd} a decline in average variance of GPU power draw with power-capped jobs. These changes---decreases in temperature and power draw across the distribution along with more stable behavior in temperature and power draw fluctuations (i.e., less variance)---under GPU power-capping may hold implications for prolonging the lifespan of GPUs and sustainable hardware strategies in the operational management of HPC/data-centers.

\subsection{Hypothesis Testing \& Inference}
We also attempt to quantify the effect size of reduced GPU temperature and power draw more rigorously. The empirical distributions show support for the hypothesis that, from the samples collected for our analysis, GPU power capping is associated with lower GPU temperatures and lower GPU power draw on average. To help determine how significant these temperature/power differences are between power-capped and un-capped jobs and how likely these conclusions extend beyond our data sample, we use hypothesis testing to assess the degree to which reductions in temperature and power draw from power capping are meaningfully different from random chance. 

To do so, we employ a one-sided Welch's t-test with non-equal variance \cite{welchT} to determine the statistical significance and likelihood that there exists a difference between the means of two groups (capping vs. no capping). If power-capping does induce meaningful reductions in temperature and power draw, then we would see evidence from our data that rejects the null hypothesis of capping having no effect.

We define $\bar{Y}_1$ as the sample mean of the outcome variable for power-capped jobs and $\bar{Y}_0$ as the sample mean outcome for un-capped jobs. For instance, $\bar{Y}_1$ can be the average GPU temperature (or power draw) of power-capped jobs and $\bar{Y}_0$ the same quantity but for un-capped jobs. The null hypothesis $H_0$ is our ``no effect'' hypothesis, which represents the possibility that power-capping produces no significant effect on outcome variables (e.g., temperature) relative to un-capped jobs so that outcomes between capping vs. no capping are roughly equal. The alternative hypothesis $H_{\alpha}$ represents the possibility that outcomes under power-capping are smaller than under no capping (e.g., lower power draw, temperature). Formally, we can represent our hypothesis test as follows:
    \begin{equation}
    H_0: \bar{Y_1} = \bar{Y_0}, \ \ \ \ H_{\alpha}: \bar{Y_1} < \bar{Y_0}
    \label{eqn:hyp_test}
    \end{equation}

Put differently, given the data/sample on hand, we want to see how likely the null hypothesis is true (no difference between the capped and un-capped group so the outcomes are roughly the same) when put against our alternative hypothesis $H_{\alpha}$: that the average outcome (GPU temperature, power draw) of the power capped group is less than the un-capped. In Table \ref{tbl:stat_sig}, we see that differences in GPU power draw and temperatures are statistically significant between power-capped and uncapped jobs across the distribution: in other words, our null hypothesis of no effect between capped and un-capped jobs is likely false (i.e., the probability of observing the data we do if the null hypothesis---no significant difference between capped and uncapped jobs---were true is less than 0.01\%). As such, the observed decrease in both GPU power draw and temperature across the entire distribution in power-capped jobs are highly significant, statistically and economically.

\begin{table}[htbp!]
\centering
\small
\begin{tabular}{@{}llccc@{}}
\toprule
Variable &  & $\alpha=1\%$ & $\alpha=0.1\%$ & $\alpha=0.01\%$ \\ \midrule
GPU Power Draw (Mean) &  &  \checkmark   &  \checkmark     &  \checkmark \\
GPU Power Draw (50th) &  &  \checkmark   &  \checkmark     &  \checkmark      \\
GPU Power Draw (90th)
         &  &    \checkmark   &  \checkmark     &  \checkmark   \\ \midrule
GPU Temperature (Mean)
         &  &  \checkmark   &  \checkmark     &  \checkmark       \\ 
GPU Temperature (50th)
         &  &  \checkmark   &  \checkmark     &  \checkmark       \\ 
GPU Temperature (90th)
         &  &  \checkmark   &  \checkmark     &  \checkmark   \\ \bottomrule         
\end{tabular}
\caption{\small \textbf{Results from a one-sided Welch's t-test with non-equal variance and Mann-Whitney U-test, testing whether power capped jobs have lower temperature and power draw than uncapped jobs on average, at the 50th percentile, and the 90th percentile under different significance levels $\alpha$: 1\%, 0.1\%, and 0.01\%.} Differences in temperature and power draw between power-capped and un-capped jobs are highly and statistically significant with p-values virtually at zero. The null hypothesis of no effect (i.e., no difference in average outcomes between capped and uncapped jobs) can be rejected with high likelihood---in other words, the alternative hypothesis that power capping is associated with power draw and temperature is very likely to be true even beyond our sample.}
\label{tbl:stat_sig}
\end{table}

\subsection{Average Treatment Effect (ATE) Estimation}
Hypothesis testing can be useful, but several issues can still arise. First, because GPU power-capping was not instituted in a randomly balanced way, potential selection bias can skew our analysis. Secondly, although our hypothesis testing has detected a significant difference between the two groups, it does not fully address the question of how much of said effect is attributable to, or caused by, power capping itself. To address these issues, we attempt to estimate the average treatment effect of power capping on GPU temperatures and power draw via a causal inference approach that tries to mitigate potential problems with bias and attribution.

One way to estimate the effects of power capping on outcome variables like GPU power draw and GPU temperature is through calculating an \textit{average treatment effect} (ATE). The ATE measures the average difference in outcomes between groups/observations that have received some treatment (e.g., medication, policy change). The estimation of ATEs sees widespread usage in evaluating policies, medical therapies, product launches, and more. Here, we can consider the treatment to be power capping; some jobs with GPUs are power-capped (i.e., receiving the treatment) while others are not. Within this framework, we may be interested in the ATE of power capping GPUs on GPU temperatures and power draw---in both the direction of the effect as well as its size and statistical significance. Ideally, we hope that power capping will decrease temperatures and power draw as a means of improving energy sustainability. If the ATE is properly estimated, negative, and statistically robust, then we can more confidently conclude that the act of power capping induces a decrease in temperature and power draw.

More formally, we define $Y_1 = (Y | T = 1)$ to be the response of the treated group. Equivalently, $Y_1$ is the response or outcome $Y$ of the group that has received treatment $T \in \{0,1\}$ where $T=1$ corresponds to having received treatment/power capping and $T=0$ corresponds to no treatment/capping. We can similarly define $Y_0 = (Y | T = 0)$ to be the outcome of the group that did not receive treatment. Finally, we can formally express the average treatment effect as $ATE = \mathbb{E}(Y|T=1) - \mathbb{E}(Y|T=0)$ where $\mathbb{E}(\cdot)$ is the expectation operator. In other words, we are interested in the expected or average effect of the treatment by comparing the outcome of the treated group with that of the untreated. 

One common way \cite{causalinf} to estimate this is through linear regression or ordinary least squares (OLS):
    \begin{equation}
    Y = \hat{\beta}_0 + \hat{\beta}_1 T + \varepsilon  
    \label{eqn:ate_ols}
    \end{equation}

where $Y$ is our response variable of interest (e.g., GPU power draw or temperature),  $T \in \{0,1\}$ is a binary variable indicating whether an observation received treatment (i.e., power-capped), with one corresponding to having treatment and zero to no treatment, and $\varepsilon$ is a nuisance parameter that accounts for noise or other variation not accounted for by our model. Our parameter of interest $\hat{\beta}_1$ will be our estimate of the ATE of power-capping on our response $Y$. To see this, note that $ATE = \mathbb{E}(Y|T=1) - \mathbb{E}(Y|T=0) = \hat{\beta}_1$ where the first equality follows from the definition of the ATE and the second equality follows from straight-forward calculation by substituting Eq. \ref{eqn:ate_ols} into the definition. 

From the results in Table \ref{tbl:ate}, we see that a direct estimate of the effect of power-capping on GPU power draw and temperatures are not only statistically significant from the p-values, but also economically significant: power-capping can reduce GPU power draw by $10.1$ W and GPU temperature by about $4.8$ degrees Celsius on average. 

\begin{table}[htbp!]
\centering
\small
\begin{tabular}{@{}lccc@{}}
\toprule
Response & ATE & p-value & $95\%$ CI \\ \midrule
GPU Power Draw (Watts)   &  $-10.10$   & $0.00$   & $[-10.77, -9.44]$ \\
GPU Temperature ($^{\circ}\text{C}$)  & $-4.78$    & $0.00$    &
$[-4.93, -4.62]$\\
\bottomrule
\end{tabular}
 \caption{\small \textbf{The average treatment effect (ATE) of power capping on average GPU power draw and temperature along with their associated p-values and 95\% confidence intervals.} A direct estimate of the treatment effect of power-capping shows a size-able and statistically significant effect on reducing GPU power draw and temperature---reducing per job's power draw by about 10 W and 4.8 degrees Celsius on average. Values are estimated via OLS.}
\label{tbl:ate}
\end{table}

In Table \ref{tbl:ate_efficient}, we repeat the same estimation procedure but only for ``efficient'' jobs---which we define as those that have an average GPU utilization above $70\%$. In other words, these jobs are characterized as ones that use GPUs efficiently by utilizing as much of the GPU as possible. The goal is to see if these efficient jobs see different impacts from capping relative to the capped cohort. Indeed, we see that the direct estimation of these effects also show decreases in power draw and temperature for GPUs for the average efficient job: about 13.3 W and 7.5 degrees Celsius on average. The effect of reduced temperature/power draw is higher for more efficient jobs likely due to the fact that jobs with higher GPU utilization will be impacted more from changes to the GPU such as power capping.

\begin{table}[htbp!]
\centering
\small
\begin{tabular}{@{}lccc@{}}
\toprule
Response & ATE & p-value & $95\%$ CI \\ \midrule
GPU Power Draw (Watts)   &  $-13.30$   & $0.000$   & $[-14.61, -12.00]$ \\
GPU Temperature ($^{\circ}\text{C}$)  & $-7.48$    & $0.000$    &
$[-7.80, -7.17]$\\
\bottomrule
\end{tabular}
 \caption{\small \textbf{The average treatment effect (ATE) of power capping on average GPU power draw and temperature for ``efficient'' jobs (defined as GPU utilization greater than $70\%$) along with their associated p-values and 95\% confidence intervals.} A direct estimate of the treatment effect of power-capping shows a sizeable and statistically significant effect on reducing GPU power draw and temperature---reducing per job's power draw by about -13 Watts and 7.5 degrees Celsius on average. Values are estimated via OLS.}
\label{tbl:ate_efficient}
\end{table}

However, since assignment of power-caps was not distributed randomly across jobs and GPUs, there is the risk of selection bias and other confounding effects which may mis-characterize the true effect of GPU power-capping on GPU temperatures and power draw---as might be the case with observational data. When the treatment is not randomly assigned, estimating a causal effect requires caution especially around potentially confounding variables. We attempt to mitigate these biases in our estimates in the next section.

\subsection{Bias Mitigation \& Matching for ATE}
To try and mitigate the confounding effect of non-random assignment of GPU power-capping in estimating its causal effect on GPU temperatures and power draw, we also estimate the ATE through matching and bias-adjusted matching \cite{causalinf}. Matching attempts to address the potential biases of non-random assignment through finding groups within the data that contain both control and treatment group members. These smaller groups are formed from observations that are very similar to one another (i.e., similar covariate/feature values) with the exception that some received treatment and others do not. The idea is that by grouping observations together this way, we can implicitly control for confounding biases and instead focus on estimating the effect of the treatment. Further bias-adjustments can also be made atop the matching estimator for ATE to help reduce bias further. 

More formally, we define our estimate of the average treatment effect ($\hat{ATE}$) through matching as:
   \begin{equation}
    \hat{ATE} = \frac{1}{N} \sum_{i=1}^N (2T_i - 1)(Y_i - Y_{j,m}(i))
    \label{eqn:ate_match}
    \end{equation}
where $T_i \in \{0,1\}$ again indicates the treatment of observation $i$, $N$ is the total number of observations in our sample, $Y_i$ is the response of observation $i$ (i.e., GPU power draw or GPU temperature), and $Y_{j,m}(i)$ represents the sample or observation from the other treatment group $m$ that is most similar to $Y_i$ from the matching. To find the closest matches between observations in treatment and those in control, we employ a $k$-nearest neighbors approach with $k=1$ and Euclidean distance. We also estimate a bias-adjusted variant of the matching ATE. From Table \ref{tbl:more_ate}, we see that both estimates of the ATE of GPU power-capping on GPU temperatures and power draw still indicate that power-capping produces sizable reductions in both power draw and temperature.

\begin{table}[htbp!]
\centering
\small
\begin{tabular}{@{}lccc@{}}
\toprule
Response & ATE (Matching) & Bias-Adj. ATE \\ \midrule
GPU Power Draw (Watts)    &   $-4.56$  &  $-4.71$    \\
GPU Temperature ($^{\circ}\text{C}$)   &   $-2.60$   &   $-2.04$ \\
\bottomrule
\end{tabular}
 \caption{\small \textbf{The average treatment effect (ATE) of power capping on average GPU power draw and temperature after matching and bias-adjustments.} Even after adjusting our estimates of ATE in light of potential bias stemming from non-random assignment of power-caps, we see that power caps still induce a size-able effect on reducing GPU power draw and temperature---a reduction of about 4.6 to 4.7 Watts and 2 to 2.6 degrees Celsius on average. Values are estimated via the matching estimator and its bias-adjusted variant (Eq. \ref{eqn:ate_match}).}
\label{tbl:more_ate}
\end{table}

\begin{table}[htbp!]
\centering
\small
\begin{tabular}{@{}lccc@{}}
\toprule
Response & ATE (Matching) & Bias-Adj. ATE \\ \midrule
GPU Power Draw (Watts)    &   $-10.43$  &  $-11.98$    \\
GPU Temperature ($^{\circ}\text{C}$)   &   $-5.02$   &   $-2.73$ \\
\bottomrule
\end{tabular}
 \caption{\small The average treatment effect (ATE) of power capping on average GPU power draw and temperature for ``efficient'' jobs (defined as jobs with average GPU utilization greater than 70$\%$) after matching and with bias-adjustments. Even after adjusting our estimates of ATE in light of potential bias stemming from non-random assignment of power-caps, we see that power caps still induce a size-able effect on reducing GPU power draw and temperature---a reduction of about 10.4 to 12 Watts in power draw and about 2.7 to 5 degrees Celsius on average. Values are estimated via the matching ATE estimator and its bias-adjusted variant (Eq. \ref{eqn:ate_match}).}
\label{tbl:more_ate_efficient}
\end{table}

We also repeat our matching and bias-corrected matching ATE estimates on efficient jobs to examine if power-capping affected different types of jobs un-evenly. We implement the same procedure as before but, in an attempt to estimate the effects of power-capping on more efficient jobs, we do so on jobs whose GPU utilization exceed $70\%$. From Table \ref{tbl:more_ate_efficient}, we see that the effects are slightly more pronounced; a reduction of about 10 to 12 Watts and 2.7 to 5 degrees on average. 

\subsection{Performance Impact}

Having analyzed the effects of GPU power-capping on GPU power draw and temperatures on the job-level, we briefly examine the impact of power-capping on job performance. In analyzing performance vs. energy trade-offs, we consider a power-cap to be optimal if it reduces energy by at least $10\%$ but limits performance impact to single digits (i.e., $< 10 \%$).

\subsubsection{Model Training}
\begin{figure}[htbp!]
    \centering
    \includegraphics[width=0.45\textwidth]{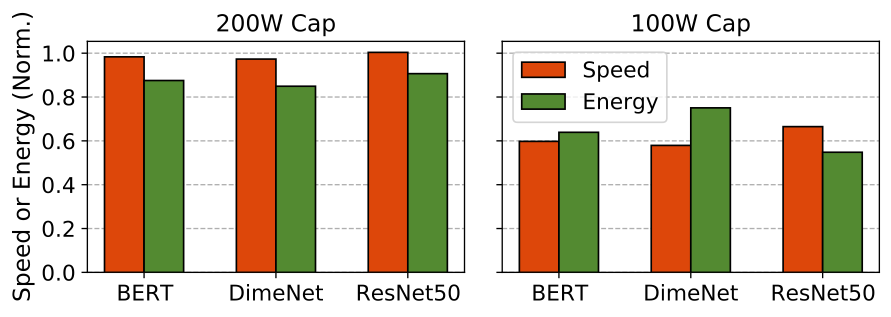}
    \caption{\small  Optimal power-capping GPUs can decrease energy expenditure with minimal adverse impact on training speed. Stricter power caps (100W) can further reduce energy but disproportionately degrades training speed. Speed and energy values are normalized to training speed and energy without power capping (e.g., a value of 0.8 corresponds to a 20$\%$ decrease in speed/energy relative to no power caps).}
    \label{fig:perf_impact}
\end{figure}

We first examine power capping impacts on the performance of deep learning training jobs where performance is measured in terms of speed or how quickly training completes (i.e., the inverse of runtime; longer runtime translates into slower speed and vice versa). These training jobs are not only common, but almost always require GPUs as part of training. In Figure \ref{fig:perf_impact}, we show the effects of power-capping on training speed and energy for three models (relative to no capping)---BERT (transformer-based encoder), ResNet50 (convolutional neural network), and DimeNet (graph neural network)---representative of three common domains of deep learning: NLP, computer vision, and graph learning, respectively. 

Overall, we see that power caps can produce considerable reductions in average training energy consumed for a variety of deep learning model architectures/domains under consideration with little to no impact on performance (i.e., training speed). Under the 200W cap, we see at least a 10$\%$ reduction in energy across the board but only single-digit adverse impacts ($< 5 \%$) on training speed with training speeds close to speeds under no caps (values of 1.0).

More specifically, for BERT training under a 200W cap, we see approximately a 15$\%$ reduction in energy but barely any degradation in training speed. We also see similar results for our graph neural network, DimeNet, and convolutional neural network, ResNet50: the former sees about a 15-20$\%$ decrease in energy and the latter sees about a 10$\%$ decrease. In either case, we also see negligible adverse impact on training speed relative to training speed without power capping. 

Although stricter power caps can naturally reduce energy expenditure further, they produce worse energy vs. performance trade-offs. In Figure \ref{fig:perf_impact}, we see that under the stricter cap of 100W, training speed exhibits noticeable degradation across all three models/domains. While we see about a larger 40$\%$ to 60$\%$ reduction in energy relative to no capping, we also see a 30$\%$ to 40$\%$ reduction in training speed across the board. In terms of finding a suitable or optimal power cap that best balances performance against energy savings, we see that there exist ``sweet spots'' that can produce relatively large energy savings with minimal impact on job performance as measured by training speed.

\subsubsection{Model Inference} Though less commonly studied, inference is receiving more attention due to how large language models (LLMs) conduct inference. Previous deep learning model architectures typically perform inference via a single pass of inputs: a batch of inputs go through the model once to produce a batch of corresponding outputs. However, LLMs generate output auto-regressively where each preceding token is fed back into the model to produce the next token. This semi-recursive property lends an additional layer of complexity to LLMs in addition to their size, making them an interesting subject of study.

\begin{table}[htbp!]
    \centering
    \small
    \begin{tabular}{lcccccc}
    \toprule
        Output & \multicolumn{2}{c}{Speed} && \multicolumn{2}{c}{Energy} \\
        length & \multicolumn{2}{c}{(rel. to no cap)} && \multicolumn{2}{c}{(rel. to no cap)}  \\
    \midrule
         & 175W & 150W & & 175W & 150W \\ 
         \midrule
         256  & 0.948 & 0.847  & & 0.782 & 0.672 \\
         512  & 0.935 & 0.783 & & 0.761 & 0.653 \\
         1024 & 0.926 & 0.782 & & 0.761 & 0.654 \\
    \bottomrule \\
    \end{tabular}
    \caption{\small Effects of GPU power capping on LLaMA 65B inference: This table shows the relative performance of the LLaMA 65B model on output lengths of 256, 512, 1024 using four NVIDIA A100 GPUs. Results are shown for power caps at 175W/150W and are relative to performance without power caps (i.e., no caps $= 1.0$).}
    \label{tab:inference-power-cap}
\end{table}
As such, we study LLaMA 65B \cite{touvron2023llama}, the largest variant of one of Meta AI's LLMs and how power-capping can impact inference performance/energy. Given the significant memory requirements for this LLM, we use this opportunity to examine the potential effects  with our limited number of A100s in contrast to the V100s used for our earlier analyses. In Table \ref{tab:inference-power-cap}, we see results similar to those on the V100s for model training; capping at 175W produces a good trade-off that reduces inference speed by only 5-8$\%$ but can reduce energy by about 22-24$\%$ across all output lengths. Stricter capping at 150W reduces energy further, but by a diminishing amount while also resulting in noticeable speed degradation.

\section{Conclusion}
We presented an analysis of the effects of power capping GPUs at an HPC/datacenter scale. Our analysis shows a reduction in GPU operating temperature and GPU energy usage attributable to power capping. These reductions may hold the potential to improve hardware reliability and increase the mean time between failures; by increasing hardware life, datacenter operators have the potential to reduce the embodied carbon costs of hardware manufacturing and deployment. In addition, providing HPC and cloud users the ability to control the GPU power cap has the potential to enable AI researchers and developers to actively make a choice to reduce energy used for AI compute. In future work, we hope to release more detailed analysis of even more data to study the effects of power capping on power usage effectiveness (PUE) and the longer-term impacts, behavioral and otherwise, of power capping alongside additional factors like variations external temperature and seasonality, weather, system utilization, and more. 

Several important questions nonetheless remain. For instance, how do differences between academic HPC and commercial cloud provider workloads impact the implementation or effects of power-capping? On the system level, what job-scheduler or multiplexing challenges exist for GPU power-capping? With current GPU shortages, if power-capping can help extend hardware lifetime, can commercial cloud providers get more out of existing/older or heterogeneous hardware? What savings, if any, are there in using a GPU with an extended life-time due to power-capping versus purchasing a new GPU?
Another avenue that can build off of our work is the development of dynamically adaptive GPU power caps that can adjust power caps for scheduled jobs based on different workload characteristics. An adaptive or variable power-capping system may further improve the energy efficiency of HPC/datacenters by finding more precise or optimal efficiency vs. performance trade-offs. We hope our future work will continue to inform better strategies to improve the sustainability of hardware accelerators and other resources required by AI development. \\

\noindent
\textbf{Acknowledgments:} Research was sponsored by the United States Air Force Research Laboratory and the Department of the Air Force Artificial Intelligence Accelerator and was accomplished under Cooperative Agreement Number FA8750-19-2-1000. The views and conclusions contained in this document are those of the authors and should not be interpreted as representing the official policies, either expressed or implied, of the Department of the Air Force or the U.S. Government. The U.S. Government is authorized to reproduce and distribute reprints for Government purposes notwithstanding any copyright notation herein. The authors acknowledge the MIT SuperCloud team: William Arcand, David Bestor, William Bergeron, Chansup Byun, Daniel Burrill, Matthew Hubbell, Michael Houle, Hayden Jananthan, Mike Jones, Jeremy Kepner, Anna Klein, Guillermo Morales, Peter Michaleas, Lauren Milechin, Julie Mullen, Andrew Prout, Albert Reuther, Antonio Rosa, and Charles Yee.

\balance
\bibliographystyle{IEEEtran}
\bibliography{IEEEabrv,bibliography}

\end{document}